\documentclass[12pt]{article}
\usepackage[pctex32]{graphics}
\begin{document}
\begin{center}
  
{\Large \bf Particle Creation in Kaluza-Klein Cosmology}

\vspace{1cm}

                      Wung-Hong Huang\\
                       Department of Physics\\
                       National Cheng Kung University\\
                       Tainan,70101,Taiwan\\

\end{center}
\vspace{2cm}

\begin{center} {\large \bf ABSTRACT }\end{center}

   We exactly calculate the particle number $N$ of scalar fields which are created from an initial vacuum in certain higher-dimensional cosmological models. The spacetimes  in these models are the four-dimensional Chitre-Hartle or radiation-dominated  universe with extra spaces which are static or power-law contracting. Except for some  models in which no particles could be produced, the distribution of created particles  shows a thermal behavior, at least in the limit of high three-dimensional "momentum"  $k$. In some models, $N$ does not depend on the magnitude of the extra-dimensional  "momentum" $k_c$ if $k_c$ is nonvanishing.  A cutoff momentum $k_c$ may  emerge in some models, and particles with $k\le k_c$ could not be produced. We also  discuss these results.

\vspace{4cm}
\begin{flushleft}
E-mail:  whhwung@mail.ncku.edu.tw\\
Publihed in Physics Letters A140 (1989) 280-284
\end{flushleft}


\newpage

\section  {Introduction}
   Particle production on the background geometry of cosmology has been studied by  several authors [1,2]. To determine the number of produced particles one must  calculate the overlap function between the initial and final state of the particle, which
is induced by the dynamical changing of the universe. Many exactly solvable models  have been found for theories in a 3-flat Robertson-Walker universe [2].  However, it  is difficult to solve exactly anisotropic models. To our knowledge, only the exact  solution for the Kasner-like model in three dimensions has been obtained [3]. This  then led Birrell and Davies [4] (see also ret. [5]) to develop a perturbation method for  solving it. As the method they exploited could only treat models with small  anisotropies, the problem of particle production in "realistic" higher-
dimensional unified theories, e.g. Kaluza-Klein supergravity [6], superstring theory  [7], etc., in which three spaces are expanding while the other extra spaces are  contracting, could not be studied. Thus it is useful to evaluate exactly the number of  particles produced on a Kaluza-Klein cosmology.

   In this paper we will calculate exactly the particle number of scalar fields which are  created from an initial vacuum in certain Kaluza-Klein cosmologies. The spacetimes  in these models are the four-dimensional Chitre-Hartle or radiation-dominated  universe with extra spaces which are static or power-law contracting. The results  show that no particle could be created from some spacetimes even though they
are not conformally flat. In some models, N does not depend on the magnitude of the  extra-dimensional "momentum" $k$ if $k\ne 0$, and no particle could be
produced in the modes in which the three-dimensional momentum $k\le k_c$. (The cutoff momentum $k_c$ depends on the model.) We also find that the distribution of created particles does show a thermal spectrum at sufficiently short wavelengths, in agreement with conventional belief [8,9].  

   In section 2 we briefly derive the formulas and we use them to evaluate the number of particles created in some five-dimensional models in section 3. The extension to higher-dimensional theories and some discussions are given in section 4. 

\section {Cosmological Particle Creation: Formulation}

   The models considered in this paper are a scalar field $\Psi$ minimally coupled to a five-dimensional spacetime with metric $g_{\mu\nu}$.   The scalar field will satisfy the wave equation

$${1\over \sqrt g}\partial_\mu (\sqrt g g^{\mu\nu} \partial _\nu \Psi) + m^2\Psi = 0.  \eqno{(2.1)}$$
\\
The line element we used is

$$ds^2 = dt^2 - A^2(t) d{\bf x}^2 - B^2(t)dx_5^2, \eqno{(2.2)}$$
\\
where $dx_5$ is the coordinate of the fifth space. Introducing the so-called conformal time $\eta$ defined by 
$$d\eta = A(t)^{-1} dt ,            \eqno{(2.3)}$$
we write eq. (2.2) as

$$ds^2 = A^2(t)(d\eta^2 - d{\bf x}^2) - B^2(\eta)dx_5^2.       \eqno{(2.4)}$$
\\
We now expand $\Psi$ in terms of a complete set of  modes $u_{{\bf k},k_{5}}$,

$$\Psi({\bf x}, x_5, \eta) = \sum_{{\bf k},k_{5}}  a_{{\bf k},k_{5}} u_{{\bf k},k_{5}} ({\bf x}, x_5, \eta)+   a^\dag_{{\bf k},k_{5}} u^*_{{\bf k},k_{5}}({\bf x}, x_5, \eta) , \eqno{(2.5)}$$
\\
where the creation operators $a^\dag_{{\bf k},k_{5}}$, and annihilation operators $a_{{\bf k},k_{5}}$ satisfy the usual commutation relations. After defining

$$ u_{{\bf k},k_{5}} = S^{-1}(\eta) exp[-i (k\dot x + k_5 x_5)] \Psi_{{\bf k},k_{5}}(\eta),    \eqno{(2.6)}$$
\\
where

$$S(\eta) = A(\eta) B^{1/2}(\eta),           \eqno{(2.7)}$$
\\
one can from eq. (2.1) obtain an ordinary differential equation for $\Psi_l$  ($l$ stands for ${{\bf k},k_{5}}$),

$${d^2 \Psi_l \over d\eta^2} + \Omega^2(\eta)  \Psi_l(\eta) = 0 , \eqno{(2.8)}$$
\\
where

$$\Omega^2(\eta) = k^2 + A(\eta) [k^2_5/B(\eta)^2 + m^2] -{d^2S/d\eta^2\over S} .  \eqno{(2.9)}$$
\\
Suppose the spacetime to be static at $\eta \rightarrow \pm\infty$, or so slowly varying that an adiabatic definition of the vacuum state makes sense. We can then use the WKB approximation to find the solutions of eq. (2.8) in the limits $\eta \rightarrow \pm\infty$. Taking this as the "asymptotical condition" we in general have two solutions $\Psi^{in}_l$ and $\Psi^{out}_l$  in eq. (2.8).  As both sets are complete,
we can expand $\Psi^{out}_l$ in terms of $\Psi^{in}_l$,

$$\Psi^{out}_l = \alpha_l \Psi^{in}_l +  \beta _l  \Psi^{in}_l*, \eqno{(2.10)}$$
\\
and the vacuum of the $\Psi^{out}_l$ mode will contain $|\beta_l|^2$ particles in the $\Psi^in_l$ mode [1,2]. Thus the number of produced particles $N_l$ is
$$ N_i = |\beta_l|^2.  \eqno{(2.11)}$$
Note that the Wronskian condition which guarantees the conservation of particle flux and the unitarity of the S matrix implies
$$|\alpha_l|^2 - |\beta _l|^2 = 1. \eqno{(2.12)}$$
The remaining work is to find the exact solution ofeq. (2.8) for a given metric, which in turn exactly determines the value of  $\beta _l$ and thus the number $N_l$. Before doing this, however, we will briefly discuss two classes of second order differential equations which correspond to the wave equations (2.8) for the models analyzed in the next section. 

$$ (a) ~~~~~d^2\Psi /d\eta^2 + (F e^{\lambda\eta} + G) \Psi=  0.~~~~~ \lambda >0.    \eqno{(2.13)}$$
\\
The general solutions can be expressed in terms of  the Bessel function,

$$ \Psi (\eta) =  C Z_\nu (X) ,~~~  X= {2\sqrt F\over \lambda} e^{\lambda\eta/2} ,\eqno{(2.14)}$$
\\
where C is a normalization constant and $\nu$  is the order of the Bessel function,
$$ \nu = - {2 i \sqrt G\over \lambda}.\eqno{(2.15)}$$
\\
At the initial time, $ t \rightarrow - \infty$ , $\Omega ^2$ approachs the constant $G$, and the state is static. On the other hand, at the final time, $ t \rightarrow + \infty$ , $\Omega ^2$ does not become a constant number and the physical state is not static. However, as the measure of adiabaticity $\Omega ^2 d\Omega /d\eta \rightarrow 0$ , we can use the WKB method to define the adiabatic vacuum. Using these as the asymptotical conditions we can find the in-mode and out-mode solutions:

$$\Psi^{in}_l \propto J_\nu (X),  \eqno{(2.16a)}$$
$$\Psi^{out}_l \propto H_\nu^{(2)}(X)  , \eqno{(2.16b)}$$
\\
where $J_\nu (X)$  and $H_\nu^{(2)}(X)$ are the Bessel function of the first class and the Hankel function, respectively.   Note that we cannot choose the initial state at $\eta =0$, as the measure of adibaticity at $\eta =0$ is not zero and the adiabatic definition of particles does not make sense. 

   Using the relation

$$H_\nu (X) = - {i\over sin\pi \nu} [e^{i\pi \nu}J_\nu(X) - J_{_\nu}(X)],    \eqno{(2.17)}$$
\\
one finally obtains the number of produced particles, 
$$ N= {1\over |e^{2i\pi\nu}| - 1} . \eqno{(2.18)}$$
Note that if $\nu$ is a real number then no panicles will be produced.

$$ (b) ~~~~~ d^2\Psi /d\eta^2 + (F \eta^{-2}  + G) \Psi=  0.   \eqno{(2.19)}$$
\\
This case will have asymptotically static states if the universe begins at $\eta = \pm \infty $ and ends at $\eta = \mp \infty $. The solutions are

$$\Psi^{in} = \Psi^{out} \propto {\sqrt \eta}~ H^{(2)}_\nu ({\sqrt G} \eta) ,~~~ \nu = 1/4 - F, \eqno{(2.20)}$$
\\
and thus, trivially, no particles have been created.

\section { Cosmological Particle Creation: Models and Results}

    We now present the exact number of particles produced in the following pacetimes:

     Model 1. $A(\eta) = A_0 e^{\alpha \eta}$, $B(\eta) =B_0$. This spacetime describes the four-dimensional Chitre-Hartle universe [10] with a static extra space. (Note that the Chitre-Hartle metric is a solution of the Einstein equation modified by including a one-loop quantum-gravitational correction arising from the quantum trace anomaly [11].) The wave equation to be solved is eq. (2.13) with

$$  F= A^2_0 (k^2_5 B_0^{-2} + m^2), ~~~~ G= k^2 - \alpha^2,~~~ \lambda =2\alpha.                        \eqno{(3.1)}$$

Thus the number of produced particles is

$$N_{{\bf x}, x_5} = \left [exp[(2\pi/alpha) \sqrt{k^2-\alpha^2}]-1\right]^{-1} .     \eqno{(3.2)}$$
\\
It is independent of the fifth dimensional "momentum" $k_5$ and particle mass m, and shows a black-body spectrum when $k>>\alpha$.  Note that particles with mode $ k\leq \alpha$ could not be created. One must also notice in the case $m=k_5=0$,  that $F=0$ and eq. (2.13) admits a plane wave solution, therefore, no particle could be created.

   The extra space in this model is static, while that in the next model is contracting.

   Model2.   $A(\eta) =A_0~ exp^{\alpha\eta}$, $B(\eta) =B_0~ exp^{-\beta\eta}$. This space-time describes the four-dimensional Chitre-Hartle metric with a power-law contracting extra space if  $\beta > 0$. When $m = 0$ then the wave equation to be solved is eq. (2.13) with 

 $$F= {A_0^2\over B_0^2} k^2_5, ~~~~, G = k^2 -(\alpha -1/2 \beta)^2, ~~~ \lambda =2(\alpha + \beta) .\eqno{(3.3)}$$
\\
Thus the number of produced particles is

$$N_{{\bf x}, x_5} = \left[exp[(2\pi/(\alpha+\beta) \sqrt{k^2-(\alpha-1/2 \beta)^2}]-1\right]^{-1} .     \eqno{(3.4)}$$
\\
It is independent of the fifth dimensional "momentum" $k_5$, and shows a black-body spectrum when $k>>|\alpha-1/2 \beta|$. Note that no particle with mode
$k\le|\alpha-1/2 \beta|$ could be produced. One must also notice if $k_5=0$, that $F=0$ and eq. (2.13) admits a plane wave solution, therefore, no particle could be
produced.

    Model 3. $A(\eta)=A_0 \eta$, $B(\eta) = B_0$. This spacetime describes the radiative Robertson-Walker universe with a static extra space. The wave equation to be solved is 

$$ d^2\Psi /d\eta^2 + [k^2 + A^2_0 (m^2 + k^2_5/B^2_0)] \Psi=  0.   \eqno{(3.5)}$$
\\
This equation has been analyzed by Audretsch and Schafer [12], after replacing $m^2 + k^2_5/B^2_0$ by $m^2$. As discussed in that paper, in order to have the adi-
abatic region in which the ingoing and outgoing particle definition makes sense without changing the Einstein equation for all time, the universe is completed in passing through the singularity by a time-symmetric one. Then the wave equation could be solved by the parabolic cylinder function, and the number of produced particles is

$$N_{{\bf x}, x_5} = exp\left(-\pi k/A_0 \sqrt{k_5^2 B_0^{-2} + m^2 }\right).\eqno{(3.6)}$$
\\
Our model is only a simple extension of that studied in ref. [12], and the result also shows a non-relativistic thermal spectrum.

Model 4. $A(\eta)=A_0 \eta$, $B(\eta) = B_0 \eta^{-2C}$. This spacetime describes the four-dimensional radiative universe with a contracting extra space if $C > 0$. For the case $C =1/2 $, it just describes the five-dimensional Kasner universe. If  $m = k_5=0$ then the wave equation to be solved is eq. (2.19) with

$$F = C(C-1), ~~~~ G = k^2, \eqno{(3.7)}$$
\\
and no particle will be produced eventually. Note that when $\eta \rightarrow 0$, $B(\eta) \rightarrow \infty$, so during the evolution the extra space will be infinitely large. To see that the production of no particles is not always the consequence
of having an infinite space, we consider the next model in which the exact solution could be found even for the non-zero ks mode. (However, the extra space is expanding.)

    Model 5. $A(\eta)=A_0 \eta$, $B(\eta) = B_0 \eta^2$. If  $m = 0$ then the
wave equation to be solved is just eq. (2.19) with

$$ F =A^2_0 B^{-2}_0 k^2_5, ~~~~ G = k^2 - 2,   \eqno{(3.8)}$$
\\
and no particles will be produced also.

   Model 6. For any model whose $S(\eta) (=A(\eta) B^{1/2}(\eta))$ is a power function of $\eta$, in the case $k_5 = m =0$ the wave equation becomes as eq. (2.19),
and thus no particle could be created.

\section {Discussions}

Although all the above models are described in 1+3+1 dimensions, the extension to higher dimension is straightforward. Let us consider (1 +3+M)-dimensional theories in which the scale functions of the three dimensions and extra M dimensions are denoted as $A(\eta))$ and $B(\eta)$, respectively. Then the wave equation ofeq. (2.8) could still be used if the $S(\eta)$ function is now defined by 

$$S(\eta) =A(\eta) B(\eta)^{M/2}. \eqno{(4.1)}$$
\\
One can see that all the above results will be unchanged for theories in any other dimension. 

Let us discuss our results and address some questions arisen from them.

  (1) In models 1, 2 and 3, the particle spectrum does show a thermal behavior, as a consequence of the general property argued by Parker [8] and shown recently by Kandrup [9J.

  (2) A cutoff momentum is displayed in models 1 and 2, below which no particle will be produced. This is because below the threshold the G term in eq. (2.13) is negative, and at initial time, $\eta \rightarrow -\infty$, $\Omega^2$ becomes a negative number. This implies that the wave function $\Psi$ is of exponential type, and the particle cannot propagate in the space. Thus no particle will be created.

  (3) For the massless case of models 1 and 2, one could assert that particle creation is independent of the fifth dimensional momentum $k_5$ only if $k_5$ is non-vanishing. A "thorough" independence of $k_5$ shows itself only in the massive case in model 1.  A "physical explanation" for this "independence" is unknown now.

  (4) Intuitively, particles of higher mass are produced in smaller quantity, so the fact that the particle spectrum in model 1 does not depend on the particle mass (excluding the case $k_5$, $m=0$, in which no particle is produced) is strange. This needs fur-
ther explanation.

  (5) In models 4, 5 and 6, no particles are created, although these spacetimes are not conformally flat.   Is there a simple argument against particle production in these models ? This remains to be found out. 

(6) Furthermore, according to the intuition that the entropy of a particle field should be directly connected with the number of quanta in the field, how will the Kandrup-Hu entropy [ 13,14 ] be changed in these no-particle-production models? Further analyses are needed to answer this question. 

  Work on these problems is in progress.


\newpage
{\begin{center} {\large \bf  REFERENCES} \end{center}}
\begin{enumerate}

\item   L. Parker, Phys. Rev. Lett. 21 (1968) 562; Phys. Rev. 183 (1969) 1057; Phys.Rev.D3 (1971) 346.
\item N.D. Birrell and P.C.W. Davies, Quantum fields in curved space (Cambridge Univ. Press, Cambridge, 1982), and references therein.
\item  S.A. Fulling, L. Parker and B.L'Hu, Phys. Rev. D 10 (1974) 3905;
      B.K. Berger, Phys. Rev. D 12 (1975) 368;
      D.M. Chitre and J.B. Haitle, Phys. Rev. D 16 (1977) 251.
\item  N.D. Birrell and P.C.W. Davies. J. Phys. A 13 (1980) 2109.
\item Ya.B. ZePdovich and A.A. Starobinsky, JETP Lett. 26 (1977)252.
\item  M.J. DufT, B.E.W. Nilsson and C.N. Pope, Phys. Rep. 130 (1986) 1.
\item  M.B. Green, J.H. Schwarz and E. Witten, Superstring theory (Cambridge Univ. Press, Cambridge, 1986).
\item  L. Parker, Nature 261 (1976) 20.
\item  H.E. Kandrup, Phys. Lett. B 215 (1988) 473.
\item  D.M. Chitre and J.B. Hartle, Phys. Rev. D 16 (1977) 251.
\item  M.V. Fischeni. J.B. Hartle and B.L. Hu, Phys. Rev. D 20 (1979)1757.
\item  J. Audretsch and G. Schafer, J. Phys. All (1978) 1583; Phys.Lett.A66 (1978)459.
\item  H. E. Kandrup, Class. Quantum  Grav. 3 (1986) L55; J. Math. Phys. 28 (1987) 1398; Phys. Rev. D 37  (1988) 3505; 38 (1988) 1773.
\item  B.L.Hu and H.E. Kandrup.Phys.Rev.D35 (1987) 1776.

\end{enumerate}
\end{document}